\documentclass[pra,aps,reprint,nofootinbib]{revtex4-1}
\usepackage{graphicx}
\usepackage{multirow}
\usepackage{amsmath}
\usepackage{amssymb}
\usepackage{dsfont}
\usepackage{amsthm}
\usepackage{eucal}
\usepackage{xcolor}
\usepackage{bbm}
\usepackage{bm}
\usepackage{tikz}
\usepackage{upgreek}
\usepackage[final]{hyperref} 
\hypersetup{
	colorlinks=true,       
	linkcolor=blue,          
	citecolor=red,        
	filecolor=magenta,      
	urlcolor=black         
}
\usepackage{qcircuit}
\usepackage{pifont}
\usepackage{mathrsfs}
\usepackage{diagbox}
\usepackage{color}
\usepackage{mathtools}
\usepackage{latexsym}
\usepackage{float}
\usepackage{extdash}

\theoremstyle{plain}

\theoremstyle{plain}

\usepackage{comment}

\usepackage{xpatch}
\xpatchbibdriver{misc}
  {\printtext[parens]{\usebibmacro{date}}}
  {\iffieldundef{year}
    {}
    {\printtext[parens]{\usebibmacro{date}}}}
  {}
  {\typeout{There was an error patching biblatex-ieee (specifically, ieee.bbx's @online driver)}}

\newcommand{\ket} [1] {\left\vert #1 \right\rangle}
\newcommand{\bra} [1] {\left\langle #1 \right\vert}

\newcommand{\mean}[1]{\left\langle #1 \right\rangle}

\newcommand{\bs}[1]{\boldsymbol{#1}}

\definecolor{JM}{RGB}{4,116,149}

\definecolor{josh}{RGB}{255, 87, 51}
\definecolor{nico}{RGB}{51, 87, 255}

\begin{document}
\title{Simulating realistic non-Gaussian state preparation}
\author{N. Quesada}

\author{L. G. Helt}

\author{J. Izaac}

\author{J. M. Arrazola}

\author{R. Shahrokhshahi}

\author{C. R. Myers}

\author{K. K. Sabapathy}

\affiliation{Xanadu, Toronto, Canada}

\begin{abstract}
We consider conditional photonic non-Gaussian state preparation using multimode Gaussian states and photon-number-resolving detectors in the presence of photon loss. While simulation of such state preparation is often computationally challenging, we show that obtaining the required multimode Gaussian state Fock matrix elements can be reduced to the computation of matrix functions known as loop hafnians, and develop a tailored algorithm for their calculation that is faster than previously known methods. As an example of its utility, we use our algorithm to explore the loss parameter space for three specific non-Gaussian state preparation schemes: Fock state heralding, cat state heralding, and weak cubic-phase state heralding. We confirm that these schemes are fragile with respect to photon loss, yet find that there are regions in the loss parameter space that are potentially accessible in an experimental setting which correspond to heralded states with non-zero non-Gaussianity.
\end{abstract}

\maketitle 


\section{Introduction} \label{sec:intro}
Photonic quantum information processing is a branch of quantum technologies that is aimed at harnessing the quantum properties of light at their most fundamental level~\cite{perspectives,anderson2010continuous,2009photonic,caspani2017integrated}.  In particular, there has recently been considerable theoretical and experimental progress in the field of photonic quantum computation~\cite{kok2007linear,braunstein2005quantum,weedbrook2012gaussian}. This comprises many different aspects such as large cluster-state generation in the time~\cite{menicucci2010arbitrarily,yokoyama2013ultra,yoshikawa2016invited,2DC} and frequency domains~\cite{menicucci2008one,chen2014experimental,roslund2014wavelength}, loop-based architectures~\cite{motes2014scalable,takeda2017universal,Marek:18,qi2018linear,1812.05358}, gate model algorithms~\cite{hamilton2017gaussian,rebentrost2018quantum,arrazola2018quantum,arrazola2018using}, and hybrid computation~\cite{van2011optical,andersen2015hybrid}. In addition to quantum computation, photonic quantum technologies enable promising applications in quantum sensing~\cite{pirandola2018advances} and quantum key distribution~\cite{laudenbach2018continuous}.

An important class of photonic states are Gaussian states, which can be efficiently described and analyzed mathematically \cite{weedbrook2012gaussian,serafini2017quantum}. It has been shown that an all-Gaussian optical scheme, i.e., an algorithm that takes Gaussian states as inputs and performs operations mapping Gaussian states to Gaussian states, including Gaussian measurements, can be efficiently simulated using classical computers~\cite{Bartlett2002b}. Therefore,  non-Gaussian states and operations are compulsory resources for performing universal photonic quantum computation~\cite{Lloyd1999a}. In addition, non-Gaussianity is an advantageous resource for various other quantum information processing tasks such as quantum metrology~\cite{dowling1,dowling2}, entanglement distribution~\cite{sabapathy2011robustness}, error correction~\cite{niset2009nogo}, phase estimation~\cite{adesso2009optimal}, bosonic codes~\cite{chuang1997bosonic,bosonicterhal, michael2016new, niu2018hardware,li2017cat,heeres2017implementing}, quantum communication~\cite{sabapathy2017nongaussian}, and quantum cloning \cite{ngcloning}. The experimental generation of high-quality non-Gaussian states and gates, however, remains a key challenge and is of much interest.

In this work we develop an algorithm  based on Gaussian Boson Sampling \cite{hamilton2017gaussian, kruse2018detailed} to reduce the computational resources required for simulating non-Gaussian state preparation schemes in the presence of photon loss. In particular, we focus on schemes that involve measurement of all but one of the modes of a multimode Gaussian state using photon-number-resolving (PNR) detectors \cite{sabapathy2018near,su1,su2}. This measurement induces a non-Gaussian effect on the remaining mode that can be engineered to prepare a desired target state, conditioned on observing a particular measurement pattern. Such an output state is referred to as ``heralded'' by the measurement outcome. It is this specific framework for non-Gaussian state preparation that enables development of an algorithm that can speed up calculations compared to methods relying on a system description directly in the Fock basis. To demonstrate the utility of this speed up, we compute useful figures of merit for three non-Gaussian state preparation schemes over loss parameter space. In particular, we identify parameter space regions where the output states remain non-Gaussian even after undergoing photon loss. 

The rest of this article is organized as follows. In Sec.~\ref{sec:bg} we provide a brief introduction to three figures of merit relevant to non-Gaussian state preparation schemes. In Sec.~\ref{num} we present our numerical algorithm for their simulation. We subsequently employ this algorithm to simulate three illustrative schemes---Fock state preparation, cat state preparation, and weak cubic-phase resource state preparation---for a wide range of relevant loss parameters in Sec.~\ref{sec:ex}, before concluding in Sec.~\ref{conc}. The explicit connection between the Fock matrix elements of Gaussian states and loop hafnians is presented in Appendix \ref{app:fockgaussian}.
\vspace{-0.3cm}
\section{Theoretical background}
\label{sec:bg}
In this Section we review the phase-space description of Gaussian states in terms of means and covariance matrices, and introduce figures of merit to characterize non-Gaussian state preparation.

\subsection{Wigner functions, Gaussian states, and covariance matrices}
The quantum state $\rho$ of an $\ell$-mode system can be uniquely characterized by its Wigner function \cite{serafini2017quantum}
\begin{equation}\label{Eq: Wigner}
W(\vec \alpha; \rho) = \int \frac{\text{d}\vec \xi}{\pi^{2\ell}} \text{Tr}[\rho \hat D(\vec \xi)] \exp\left(\vec \alpha^T \bm{\Omega} \  \vec \xi\right),
\end{equation}
where $\vec \alpha = (\alpha_1,\ldots, \alpha_\ell,\alpha_1^*,\ldots, \alpha_\ell^*)$ and similarly $\vec \xi = (\xi_1,\ldots, \xi_\ell,\xi_1^*,\ldots, \xi_\ell^*)$ are (bi-)vectors of complex amplitudes where the second half of the vector is the complex conjugate of the first half. The displacement operator is defined as $\hat D(\xi):=\exp(\vec{\xi}^T \bm{\Omega} \hat \zeta)$, where $\bm{\Omega}= \left[   \begin{smallmatrix}
	0 &  \mathbbm{1}_\ell \\
	-\mathbbm{1}_\ell & 0  
\end{smallmatrix} \right]$
is the symplectic form and $\hat\zeta_j$ is an operator vector of the mode creation and annihilation operators. Denoting $\ell$ as the number of modes,
we have $\hat\zeta_j=\hat a_j$ and $\hat \zeta_{\ell+j}=\hat a_j^\dagger$ 
for  $j=1,\cdots,\ell$. 
These bosonic creation and annihilation operators satisfy the canonical commutation relations $[\hat a_i, \hat a_j]$ = 0 and $[\hat a_i, \hat a_j^\dagger] = \delta_{ij}$.

A quantum state is called Gaussian if its Wigner function is Gaussian \cite{weedbrook2012gaussian}. Any multimode Gaussian state $\rho$ is completely parametrized by its first and second moments, namely the  vector of  means $\vec{\beta}$ with components 
\begin{align}\label{eq:WignerMeans}
\vec \beta_j = \text{Tr}[\rho\hat\zeta_j],
\end{align}
and the Wigner-covariance matrix $\bm{\sigma}$ with entries 
\begin{align}\label{eq:WignerCovMat}
\sigma_{jk} = \text{Tr}[\rho \{\hat\zeta_j,\hat\zeta_k^\dagger \}]/2 - \vec \beta_j \vec \beta_k^*,
\end{align}
where $\{x,y\} := xy +yx$ denotes the anti-commutator. 
In spite of the underlying infinite-dimensional Hilbert space, the Gaussian evolution of Gaussian states can be simulated efficiently because it is captured by linear transformations on the displacement vector and the covariance matrix, whose finite dimension grows only polynomially with the number of modes \cite{killoran2019strawberry}. Therefore, the full power of quantum computing is unlocked only when non-Gaussian states and operations are available~\cite{Bartlett2002b}. 

\subsection{Figures of merit \label{bench}}
Our first figure of merit is state fidelity. Given the density matrix of a prepared non-Gaussian state $\rho$, its fidelity to a pure target state $\ket{\psi}$ is given by
\begin{align}
F = \bra{\psi}\rho\ket{\psi}.
\end{align}
However, it is known that fidelity alone cannot capture all of the characteristics of a quantum state~\cite{ralph1999characterizing,mandarino2014about,mandarino2016assessing,tserkis2018simulation}. One option is to use the minimum value of the Wigner function as an indicator of genuine non-Gaussianity, as negativity of the Wigner function was shown to be a necessary condition for useful quantum computation~\cite{Bartlett2002b,veitch2012negative,veitch2013efficient,mari2012positive}. Yet Wigner negativity lacks an operational interpretation; it is not clear how it relates to the ability to perform useful tasks. Resource theories are aimed at answering precisely this question: given a specific state, how can we quantify its usefulness as a resource? For example, fault-tolerant universal quantum computation can be realized using fault-tolerant Clifford gates plus the resource of magic states \cite{gottesman1999demonstrating}. The onus of fault-tolerance is therefore the preparation of high quality magic states, which can be achieved when many copies of imperfect states are distilled to produce fewer high quality magic states \cite{veitch2014resource}. 

To quantify this distillation problem, the concept of ``mana'' was introduced as a useful monotone in the context of the resource theory of magic state distillation, which is based on the discrete Wigner representation. In the continuous-variable context, a natural analogue of mana called the Wigner logarithmic negativity (WLN) was proposed in Refs.~\cite{albarelli2018resource,quntaoqrt} as a measure of non-Gaussianity. The WLN of a single-mode state $\rho$ in terms of its Wigner function is defined as \cite{albarelli2018resource,quntaoqrt}
\begin{align}
\mathcal{W}(\rho) = \ln \left[ \int \text{d}\vec \alpha \left\vert W\left(\vec \alpha;\rho\right)\right\vert \right],
\end{align}
and was shown to satisfy the required properties of a non-Gaussianity monotone in Ref. \cite{albarelli2018resource}. 

Finally, in addition to fidelity and WLN, the success probability of producing a state is also an important consideration. Its role is two-fold. First, if one requires repeated measurements on the state for characterization, the total time $t$ required to run the experiment is given by $t = n/\left(f p\right)$, 
where $n$ is number of required runs, $f$ is the source frequency, and $p$ the probability of producing the state, i.e., the probability of observing the desired measurement outcome. Second, the probability dictates the resources required for converting any heralded scheme into a near-deterministic one. For example, in principle it is possible to make identical copies of the same experiment and collect all the outputs in such a way that a state produced from any of the copies is considered a success event. In this scenario, the required number of copies $N$ depends on the overall failure rate of the experiment $\epsilon$ and satisfies $ N > \log(\epsilon)/\log(1-p)$ \cite{sabapathy2018near}. 

\section{Simulation of heralded non-Gaussian states\label{num}}
The conditional preparation of non-Gaussian states using pure Gaussian states and PNR detectors was investigated in Ref.~\cite{sabapathy2018near} and further developed in Refs.~\cite{su1,su2} for the case of perfect transmission and unit detector efficiency. To account for experimental imperfections, losses and non-unit detector efficiencies must be included, leading to considerable simulation overhead. These imperfections are straightforward to include in the Gaussian formalism since they all can be modeled by loss channels, which map input Gaussian states to, in general, mixed Gaussian states.
Having included the imperfections in the pre-measurement Gaussian state, we develop a method to write the Fock basis representation of the non-Gaussian heralded state after measurement in the detected modes. We accomplish this 
by computing an expression for the Fock basis matrix elements $\langle \bm{m} | \rho | \bm{n} \rangle$, $\bm{n} = (n_1,\ldots, n_\ell), \bm{m} = (m_1,\ldots, m_\ell)$, of an $\ell$-mode Gaussian state $\rho$ with covariance matrix $\bm{\sigma}$ and displacement vector $\vec \beta$. 

We first define the following useful quantities:
\begin{align}
\bm{X} &=  \begin{bmatrix}
	0 &  \mathbbm{1}_\ell \\
	\mathbbm{1}_\ell & 0  
\end{bmatrix} , \\
\bm{\sigma}_Q &= \bm{\sigma} +\tfrac{1}{2} \mathbbm{1}_{2\ell},\\
T &=\frac{\exp\left(-\tfrac{1}{2} \vec \beta^\dagger \bm{\sigma}_Q^{-1} \vec \beta \right)}{ \sqrt{\text{det}(\bm{\sigma}_Q) \prod_{s=1}^\ell n_s! m_s!}}.
\end{align} 
As shown in detail in  Appendix \ref{app:fockgaussian}, the Fock matrix elements of a Gaussian state $\rho$ are given by the expression
\begin{align}\label{Eq: lhaf}
\langle \bm{m} | \rho | \bm{n} \rangle  = T \times  \text{lhaf}(\tilde{\bm{A}}),
\end{align}
where we have defined 
\begin{align}
\tilde A_{i,j} = \begin{cases}
\bar{A}_{i,j} &\text{ if } i\neq j,\\
\bar{\gamma}_{i} &\text{ if } i=j,
\end{cases}
\end{align}
and lhaf is the \emph{loop hafnian} (introduced in Ref.~\cite{quesada2018faster}), a matrix function that counts the number of perfect matchings of weighted graphs with loops. The matrix $\bar{\bm{A}}$ and the vector $\bar{\bm{\gamma}}$ are obtained from 
\begin{align}
\bm{A} &= \bm{X}	\left(\mathbbm{1}_{2\ell} - \bm{\sigma}_Q^{-1} \right), \\
\gamma^T &= \beta^\dagger \bm{\sigma}_Q^{-1},
\end{align}
as follows. For each $s=1,2,\ldots,\ell$, the column $s$ and row $s$ of $\bm{A}$ are each repeated in $ \bm{\tilde {A}}$ a total of $n_s$ times. If $n_s=0$ for some $s$, that row and column is not included. Additionally, the column $s+\ell$ and row $s+\ell$ of $\bm{A}$ are each included $m_s$ times. Therefore, since $\bm{A}$ is a $2\ell\times 2\ell$ matrix, the vector $\bm{n}$ determines which of the first $\ell$ rows and columns are kept, while $\bm{m}$ specifies which of the last $\ell$ rows and columns are kept. Similarly, $\bm{n}$ and $\bm{m}$ respectively determine which of the first and last $\ell$ entries of $\gamma$ are kept. The resulting matrix $\bar{\bm{A}}$ is a square matrix of dimension
$D = \sum_{s=1}^\ell n_s+m_s$, and the resulting vector $\bar{\gamma}$ is also of dimension $D$.

Once a Gaussian state has been specified in terms of its covariance matrix and displacement vector, Eq.~\eqref{Eq: lhaf} provides all the necessary information to compute the output state in a state preparation scheme. For simplicity, we consider the case where a Fock basis measurement is performed on the first $\ell-1$ modes and a single-mode state is heralded in the  $\ell$-th mode. The results generalize straightforwardly for any partitioning of the $\ell$ modes into detected and heralded modes. 

The unnormalized output state $\tilde{\rho}_\ell$ obtained after observing a photon pattern $\bm{n}_h=(n_1,n_2,\ldots,n_{\ell-1})$ in the detected modes is then given by
\begin{equation}
\tilde{\rho}_\ell=\sum_{n_\ell, m_\ell=0}^d \bra{\bm{n}_h,n_\ell} \rho  \ket{\bm{n}_h,m_\ell} \ket{n_\ell}\bra{m_\ell},
\end{equation}
where $d$ is a  cutoff dimension chosen for the output Hilbert space.

The normalized state $\rho_\ell$ can be recovered by computing the detection probability \begin{equation}
\tilde{p}=\text{Tr}(\tilde{\rho}_\ell),
\end{equation}
and using $\rho_\ell=\tilde{\rho}_\ell/\tilde{p}$. Note that this probability is, in principle,  dependent on the truncation $d$ used to represent the heralded state of the $\ell^{\text{th}}$ mode. However, by calculating the exact detection probability
\begin{align}
p =  \bra{\bm{n}_h} \rho_{[\ell-1]} \ket{\bm{n}_h},
\end{align}
where $\rho_{[\ell-1]} = \text{Tr}_{\ell}[\rho]$ is the reduced density matrix of the first $\ell-1$ modes, one can easily learn the truncation dimension necessary to faithfully represent the heralded state.  Once $p$ is known one simply increases $d$ up to the point where the difference between $\tilde p$ and $p$ is negligible within some numerical accuracy. 
Note that the density matrix $\rho_{[\ell-1]}$ corresponds to a Gaussian state since the partial trace of a Gaussian state is another Gaussian state \cite{weedbrook2012gaussian,serafini2017quantum}; thus $p$ can also be written in terms of a loop hafnian. 

For the purposes of this paper we are interested in the case where all but one of the occupation numbers $n_s$ and $m_s$ are fixed. As before, we fix this index to be $s=\ell$; this corresponds to preparing a single-mode state conditioned on a specific click pattern in the remaining $\ell-1$ modes. This leads to a structured matrix $\tilde{\bm{A}}$ where the rows and columns are repeated. We can then use a tailored formula derived by R. Kan \cite{kan2008moments} that allows for a faster calculation of the loop hafnian for such matrices with repeated rows and columns. 

As shown in Ref.~\cite{huh2015boson}, calculating the loop hafnian of $\tilde{\bm{A}}$ for the above case requires
\begin{align}
t =\left[\prod_{s=1}^\ell(1+n_s)(1+m_s)\right] \left[1+\frac{1}{2}\sum_{s=1}^\ell(n_s+m_s)\right]
\end{align}
operations. Defining the geometric mean 
\begin{align} G:=\left[\prod_{s=1}^\ell(1+n_s)(1+m_s)\right]^{\tfrac{1}{2\ell}}, 
\label{gm}
\end{align}
and the arithmetic mean 
\begin{align}
A:=\frac{1}{2 \ell}\sum_{s=1}^\ell[ (1+n_s)+(1+m_s)]
\end{align}
of the numbers $\{n_s+1, m_s+1\}$ for $s=1,2,\ldots,\ell$, we can rewrite the number of steps to calculate the loop hafnian as
\begin{align}\label{time1}
t=[\ell (A-1)+1] G^{2\ell}=O(\ell A G^{2\ell}).
\end{align}
Note that to prepare non-Gausian states, it is desirable that $A,G \geq 2$, as having $n_s=m_s=0$ for any $s$ is a projection onto vacuum, which is a Gaussian operation. 

The complexity of the tailored algorithm employed here can be compared with the best known algorithms for generic matrices developed by Bj\"orklund et al. \cite{bjorklund2018faster}, for which the complexity scales as $O(D^3 2^{D/2})$ with the dimension $D$ of the matrix \cite{bjorklund2018faster}, which in our case corresponds to $D = \sum_{s=1}^\ell n_s + m_s$. We can also rewrite this scaling in terms of the arithmetic mean $A$ and the number of modes $\ell$ as
\begin{align}\label{time2}
t' =O\left( (\ell A)^3 \left(\sqrt{2}^{(A-1)}\right)^{2\ell} \right).
\end{align}
By contrast, a direct simulation method based on a truncated Fock basis representation requires calculating and, importantly, storing $d^{2\ell}$ elements of the output density matrix, where $d$ is a chosen cutoff dimension. For each element, it is necessary to simulate a circuit that prepares the initial Gaussian state, the complexity of which is dominated by simulating the action of a linear interferometer. This in turn is dominated by computing the action of $O(\ell^2)$ beamsplitters, each of which requires $O(d^4)$ operations on a pair of modes in a mixed state. Therefore, the number of steps of a naive Fock basis simulation is given by 
\begin{align}\label{time3}
t'' = O(\ell^2d^4 d^{2\ell}).
\end{align} 
We note that one could also obtain all the matrix elements of the (mixed) Gaussian state using the results of Ref.~\cite{dodonov1994multidimensional} in terms of multidimensional Hermite polynomials~\cite{kok2001multi}. In this case one would have a similar exponential scaling in time and memory $\sim d^{2 \ell}$ albeit with a smaller polynomial overhead than a direct truncated Fock space simulation.

Let us now compare the number of steps in each of the three methods. 
We begin by noting that all the times considered here are exponential in $2\ell$, yet the bases in the exponential growth are quite different, namely $G, \sqrt{2}^{A-1}$, and $d$ for the tailored loop hafnian, generic loop hafnian, and truncated Fock basis methods, respectively.
Typically, the number of photons observed when measuring each detected mode is much smaller than the cutoff dimension required to accurately simulate the entire state preparation scheme, i.e., it usually holds that $G< d$. Moreover, due to the arithmetic mean-geometric mean inequality, it also holds that $2 < G \leq A$ (using Eq. \eqref{gm}), therefore implying that $t \ll t',t''$.

As shown in Ref. \cite{quesada2018faster}, for pure states one can also obtain the amplitude of the Gaussian pure state $\ket{\psi_G}$ in the multimode Fock basis $\bra{\bm{n}}$ using loop hafnians. In this case the times for the calculation of the amplitude $\langle \bm{n}|\psi_G \rangle$ using the different methods described earlier are 
\begin{subequations}
\begin{align}
t_{\text{pure}} &=O(\ell A_p G_p^\ell), \\
t'_{\text{pure}} &= O\left( (\ell A_p)^3 \left(\sqrt{2}^{(A_p-1)}\right)^{\ell} \right), \\
t''_{\text{pure}} &= O(\ell^2d^2 d^{\ell}),
\end{align}
\end{subequations}
for the tailored loop hafnian, generic loop hafnian, and truncated Fock respectively, and where the arithmetic and geometric means in this case are  denoted by $A_p:= \frac{1}{\ell}\sum_{s=1}^\ell (n_s+1)  $ and $G_p:= [\prod_{s=1}^\ell (1+n_s) ]^{1/\ell}$.

In practice, the different computational scalings in each method, both for pure and mixed states, can result in a reduction in the number of required computational steps by many orders magnitude. In the next section we will use the tailored algorithms to obtain figures of merit as a function of the loss parameters. Take as an example a plot that we will study in Sec.~\ref{sec:ex} C:  generating a single value in the loss parameter plane for the three-mode example in Fig.~\ref{fig:cubic} requires around 122 seconds on an i5 Intel quad processor at 2.2 GHz using the truncated Fock method with a cutoff of 20. Using the same hardware and the tailored loop hafnian method requires only 0.61 seconds. For this example the base coefficients are $G = 5.01$, $\sqrt{2}^{A-1} = 8.66$ and $d=20$. 

\section{Illustrative examples}
\label{sec:ex}

In this Section, we study non-Gaussian state generation involving beamsplitterss, single-mode squeezing, two-mode squeezing, and displacement operations, given by 
\begin{align}
B_{ij}\left(\theta,\phi\right)&=\exp{\left[\theta\left(e^{i\phi}\hat{a}_{i}\hat{a}_{j}^\dagger-e^{-i\phi}\hat{a}^{\dagger}_{i}\hat{a}_{j}\right)\right]},    \nonumber\\
S_{j}\left(z\right)&=\exp{\left[\tfrac{1}{2}\left(z_{j}^{*}\hat{a}_{j}^{2}-z_{j}\hat{a}_{j}^{\dagger 2}\right)\right]},\nonumber\\
S_{2}\left(\zeta\right)_{ij}&=\exp{\left(\zeta_{ij}^{*}\hat{a}_{i}\hat{a}_{j}-\zeta_{ij}\hat{a}_{i}^{\dagger}\hat{a}_{j}^{\dagger}\right)}, \nonumber\\
D_i(\alpha) &= \exp\left(\alpha \hat{a}_i^\dagger - \alpha^* \hat{a}_i\right),
\end{align}
respectively. Loss is characterized by a pure-loss channel $\mathcal{L}(\eta)$~\cite{chuang1997bosonic,ivan2011operator}, with total transmission coefficient $\eta$, i.e., the loss coefficient is $1-\eta$.

\begin{figure}[!t]
\scalebox{1.3}{\includegraphics{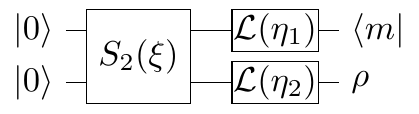}}
\caption{Gate model of a generic Fock state heralding scheme, where individual losses $\mathcal{L}(\cdot)$ are applied to each mode after the action of the two-mode squeezing operation $S_2(\cdot)$. Subsequently $m$ photons are detected in the first mode. }
\label{fig:focksf}
\end{figure}

\begin{figure*}
\begin{center}
\scalebox{0.5}{\includegraphics{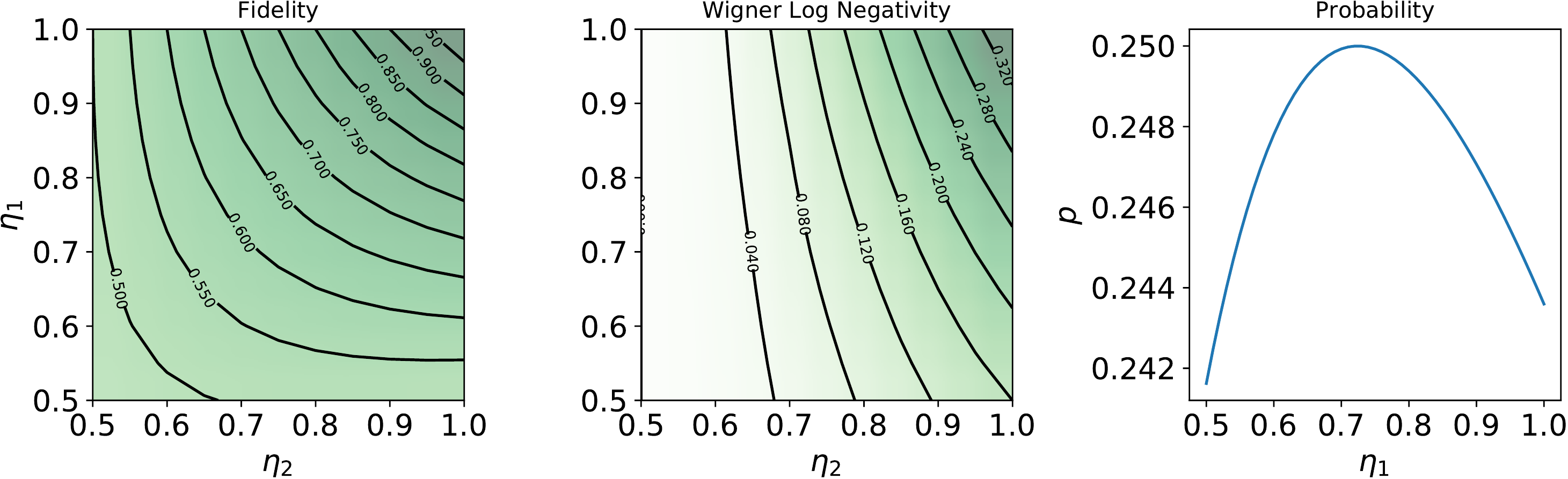}}\\
\scalebox{0.5}{\includegraphics{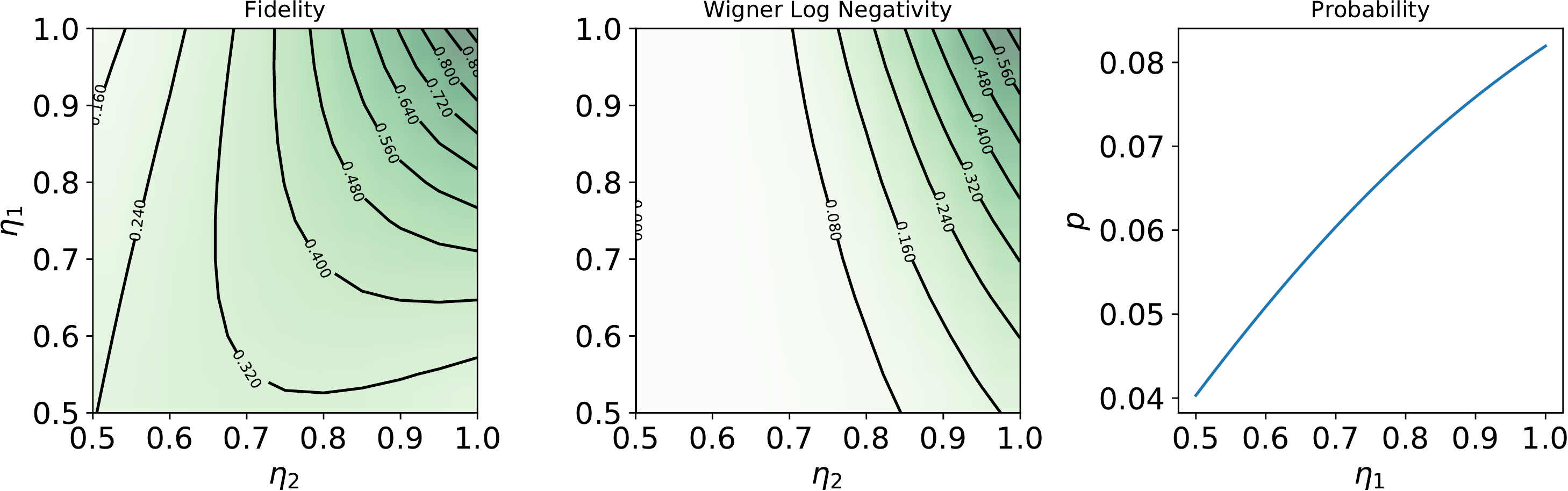}}
\end{center}
\caption{Figures of merit for the Fock state heralding scheme as a function of the transmission coefficients of each mode as depicted in Fig. \ref{fig:focksf}. We set a squeezing level of $\zeta=1.0$ (${\sim}$8.7 dB) and consider the preparation of $m=1$ (top row) and $m=3$ (bottom row) Fock states. For the preparation of single photons in the lossless case we find $p=0.243$ and $\mathcal{W} = 0.35$. Similarly, for the Fock state with $m=3$ we find $p=0.082$ and $\mathcal{W}=0.68$.}
\label{fig-fock}
\end{figure*} 

\subsection{Fock state generation from two-mode squeezed vacuum states }\label{fockexp}
Generation of single-photon states from quantum nonlinear optical processes has been considered both theoretically and experimentally~\cite{eisaman2011invited}. However, it is only more recently, with modern detector technology, that experimental focus has shifted to the heralding of Fock states containing more than one photon~\cite{ourjoumtsev2006quantum,cooper2013experimental,bouillard2019high,tiedau2019on}. Indeed, the parameter space of possible heralded states, depending on the degree of squeezing as well as various losses and efficiencies in both the detected and heralded modes, is largely unexplored. This is especially true as it concerns quantifying the utility of these states beyond their fidelity to some target state.

We consider the effect of photon loss in both the detected and heralded modes when generating Fock states from two-mode squeezed vacuum (TMSV) states. Although closed-form analytic expressions exist for some of the figures of merit discussed in Sec.~\ref{bench}  when loss is confined to the detected mode only~\cite{quesada2015very,tiedau2019on}, expressions become unwieldy once loss is considered in both modes, often requiring numerical methods. Thus, for clarity of presentation, we perform all calculations using the simulation algorithm described in Sec.~\ref{num}.

The circuit for implementing a heralded Fock state scheme is given in Fig.~\ref{fig:focksf}. Following the generation of a TMSV state, a measurement of $m$ photons in the first mode using an ideal PNR detector is said to then herald $m$ photons in the second mode. Note that losses in the detected and heralded modes may differ. For example: for TMSV states generated in OPOs~\cite{nabors1990two,uren2004efficient,neergaard2007high,morin2012high} or microring resonators~\cite{dutt2015onchip,hoff2015integrated,vernon2015strongly}, there is some finite escape efficiency $\eta_{e}$ that acts as an effective loss; the input pump light for each system is often removed with a filter $\eta_{f}$ that introduces more loss; and there may be additional loss in coupling the detected mode to fiber $\eta_{c}$ before a fiber-coupled detector with some finite detection efficiency $\eta_{d}$. Thus, identifying mode 1 of Fig.~\ref{fig:focksf} with the detected mode and mode 2 of Fig.~\ref{fig:focksf} with the heralded mode, example total transmissions could be $\eta_{1}=\eta_{e}\eta_{f}\eta_{c}\eta_{d}$ and $\eta_{2}=\eta_{e}\eta_{f}$.

In Fig.~\ref{fig-fock} we plot  the fidelity of the heralded state to an $m$ photon Fock state, the heralded state's WLN, and the success probability of detecting  $m$ photons, all as functions of $\eta_{1}$ and $\eta_{2}$. We limit our presentation here to $m=1$ and $m=3$, and a fixed level of squeezing corresponding to $\zeta=1.0$ (${\sim}8.7$ dB).  Note, however, that our custom simulation method enables quick and easy exploration of a much larger parameter space. 
Note that the detection probabilities are independent of $\eta_2$, for this value has no bearing on what reaches the PNR detector. In the absence of loss, the probability of heralding a given Fock state $\ket{m}$ for a TMSV state with real squeezing parameter $r$ is given by
\begin{align}
p_m = \frac{\mean{n}^m} {(1+\mean{n})^{m+1}}, \quad \mean{n} = \sinh^2 r,
\end{align}
and attains its maximum value $p_m^{\max}$ precisely when $\mean{n} = \sinh^2 r = m$; in the presence of loss it reaches a maximum at $m=\mean{n_{1}}=\eta_{1}\sinh^2 {r}$, which explains the local maximum in the top right ($m=1$) subplot of Fig.~\ref{fig-fock}. In general, both fidelities to $\ket{m}$ and WLNs increase as losses are reduced in both modes, though note that, in contrast to the heralding probabilities, they largely depend on $\eta_2$. As expected, fidelities approach unity as $\eta_1$ and $\eta_2$ approach unity, and WLNs are also largest as $\eta_1$ and $\eta_2$ approach unity, increasing as $m$ increases. However, we stress that the fidelity and WLN plots have different shapes, and that the very ability to explore the parameter space of the WLN with ease has been enabled by our custom simulator.

\subsection{Multi-photon subtraction and cat states}
\begin{figure}
\begin{center}
\scalebox{1.1}{\includegraphics{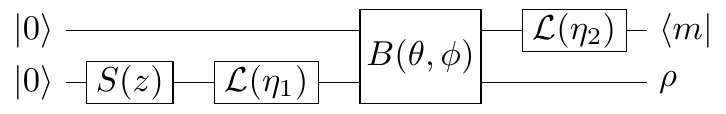}}
\end{center}
\caption{A photon subtraction schematic for the generation of cat states. A squeezed state, with squeezing parameter $z$, is generated  in mode 1 which then interacts with vacuum in mode 2 on a high-transmissivity beamsplitter. Subsequently, mode 2 is subjected to a photon number-resolving detector. Loss channels are inserted both near the beginning ($\mathcal{L}(\eta_1)$) and end ($\mathcal{L}(\eta_2)$) of the circuit. The properties of the output state are then analyzed with respect to the generation of cat states where we fix $\phi=0$, $\cos(\theta) =\sqrt{0.97}$ and $z=0.5$. }
\label{ps-cat}
\end{figure}

We now study the standard multiphoton subtraction scheme~\cite{danka1997generating,gerry1997quantum,ourjoumtsev2006generating,wakui2007photon} depicted in Fig.~\ref{ps-cat}, which includes losses at the beginning and end of the circuit. First, squeezed light is produced from a nonlinear optical source. This light is then combined on a high transmission beamsplitter (typically $97\%$ is used~\cite{gerrits2010generation}) with the vacuum mode in the other arm. Finally, a PNR measurement is performed in one mode to herald a photon-subtracted state in the other mode. Depending on the measurement outcome, states with different characteristics are heralded. 

As shown in Refs.~\cite{danka1997generating,gerry1997quantum,ourjoumtsev2006generating,wakui2007photon}, this experimental scheme is a method to prepare approximate cat states, which are superpositions of coherent states. Cat states can be either even or odd, and are defined respectively as 
\begin{align}\label{cats_def}
\ket{C_e(\alpha)} &= \frac{\ket{\alpha} + \ket{-\alpha}}{\sqrt{2(1+e^{-2|\alpha|^2})}}, \nonumber \\
\ket{C_o(\alpha)} &= \frac{ \ket{\alpha} - \ket{-\alpha}}{\sqrt{2(1-e^{-2|\alpha|^2})}}.
\end{align}
If an even (odd) number of photons is detected, then an even (odd) cat state is heralded in the other mode. 
As discussed in Refs.~\cite{danka1997generating,gerry1997quantum,ourjoumtsev2006generating,wakui2007photon},
high amounts of squeezing are required to obtain large values of $\alpha$ in the cat states.
\begin{figure*}[htb]
\begin{center}
\scalebox{0.5}{\includegraphics{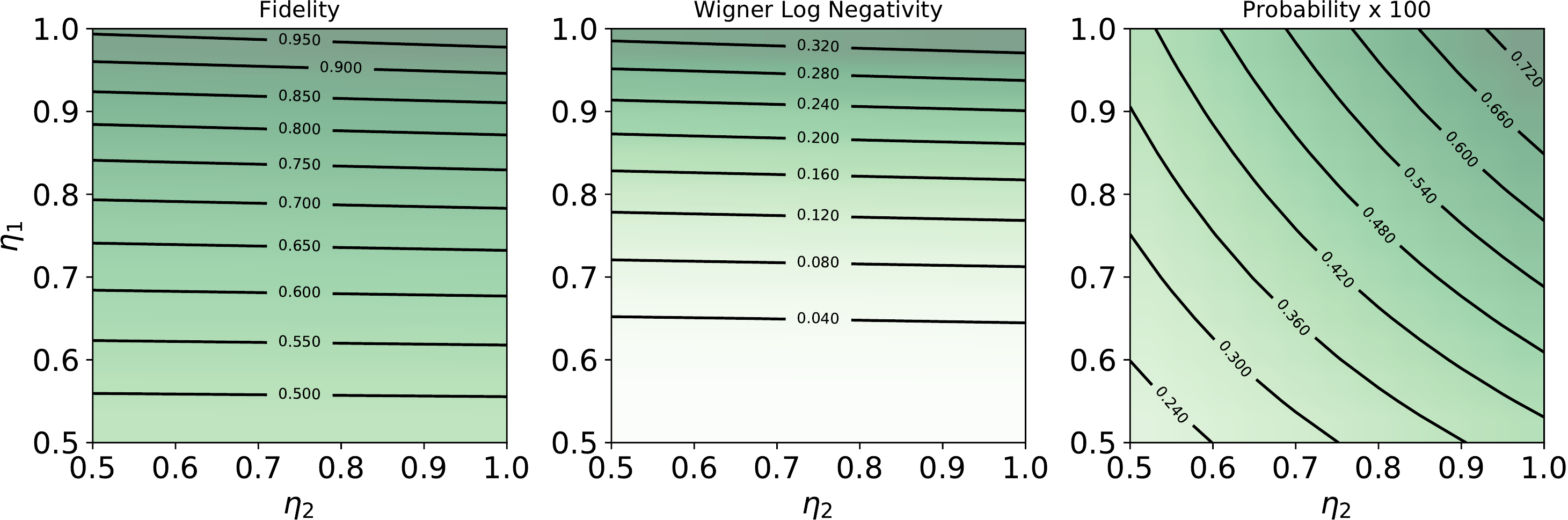}}\\
\scalebox{0.5}{\includegraphics{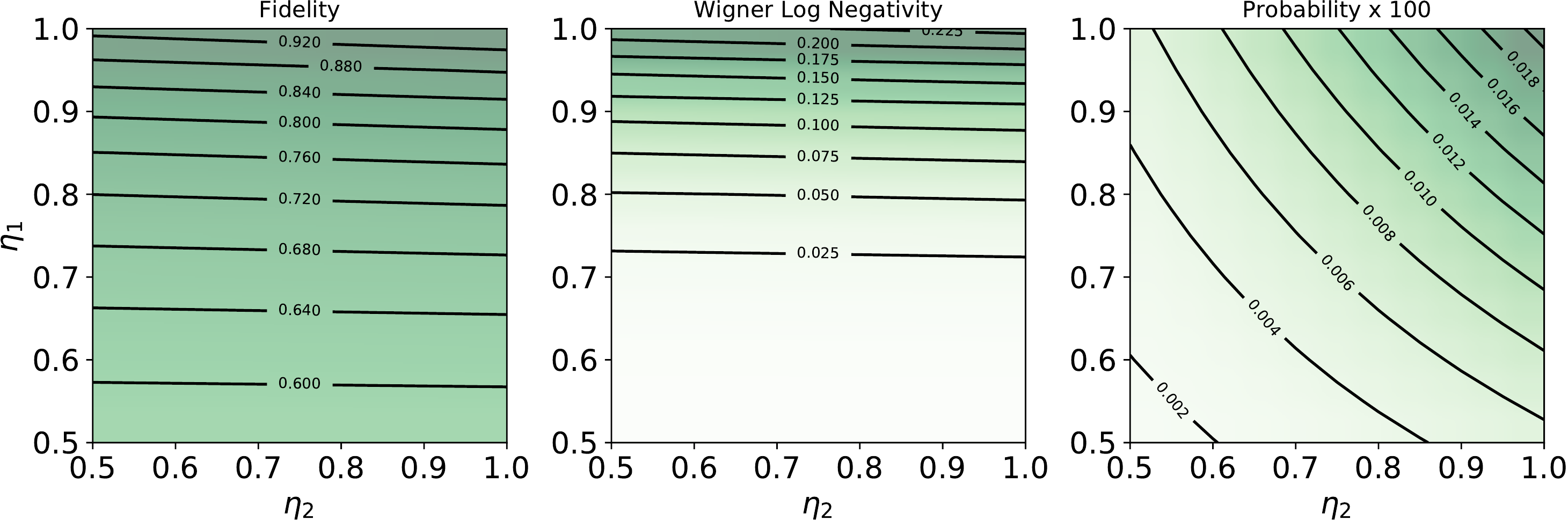}}
\end{center}
\caption{Figures of merit for the preparation of the odd cat state with $m=1$ in Fig.~\ref{ps-cat} (top row), and even cat state with $m=2$ in Fig.~\ref{ps-cat} (bottom row) as functions of the loss coefficients $\eta_1$ and $\eta_2$ in each mode. Both fidelity and WLN are largely insensitive to loss in the detected mode, but decay rapidly with loss in the output mode. Heralding probabilities are small, typically in the neighbourhood of $p\sim 10^{-4}$. The squeezing level of the input state is set to $z = 0.5$ ( $\sim$ 4.5 dB). The fidelities calculated for any value of $\eta_1$ and $\eta_2$ are taken with respect to the same lossless odd (even) cat state with $\alpha_\text{opt} \approx 1.24$ ($\alpha_\text{opt} \approx 1.33$) in Eq.~\eqref{cats_def}. For the odd cat state preparation ($m=1$), the probability, fidelity, and Wigner logarithmic negativity in the lossless case where $\eta_1 = \eta_2 = 1$ are, respectively, $p=0.0077$, $F=0.98$, and $\mathcal{W} = 0.35$. Similarly, for the even cat state $m=2$, we find $p=0.00021$, $F = 0.96$, and $\mathcal{W} = 0.23$.}
\label{fig-cat}
\end{figure*}

The fragility of cat states when subjected to photon loss can be identified in Fig.~\ref{fig-cat} where, for a given $\eta_s$ and $\eta_i$, we plot the fidelity to the lossless case ($\eta_s = \eta_i = 1$) as well as the WLN and heralding probability of the output state as functions of loss in each mode. While the probability depends on both $\eta_{1}$ and $\eta_{2}$, an interesting observation is that the fidelity and WLN are mostly insensitive to the loss in the detected mode (2). To help understand this, consider a highly transmissive beamsplitter unitary expanded as $B(\theta,0) \approx \mathbb{I}+\theta \left( a_i a_s^\dagger-a_i a_s^\dagger \right)$ for $\theta \ll 1$. Now note that for the postselection to be successful, one needs to annihilate a photon from the first mode and create one in the second mode. If the postselection was successful, it has to be that the single photon that was created in the second mode was not lost. Note that it could also be that two photons are destroyed in one mode and created in the other, with one photon being lost and the other being one measured. This is indeed a possibility consistent with the postselection employed in the protocol, but its associated probability amplitude scales like $\theta^2$, which under the assumption of a highly transmissive beamsplitter is a very small amplitude.
A similar argument can be made for the subtraction of two photons; now one needs to consider the expansion of the beamsplitter unitary up to $\theta^2$. In this case the correction to the postselection protocol due to loss when three or more photons are involved will be of order $\theta^3$, which is again a very small correction.

\subsection{Weak cubic-phase states}

\begin{figure*}
\begin{center}
\scalebox{1.3}{\includegraphics{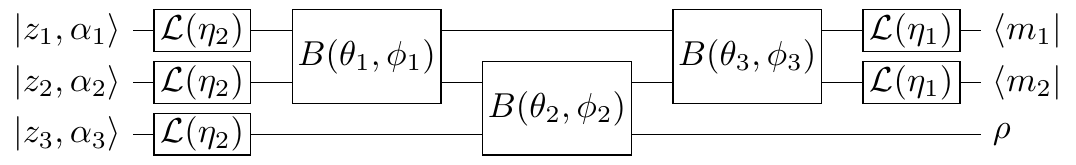}}
\end{center}
\caption{Circuit for the production of the weak cubic-phase states. The inputs are squeezed displaced states for each mode given by $\ket{z,\alpha} = D(\alpha)S(z)|0\rangle$, and the detection pattern is set to $m_1 = 1$, $m_2=2$. The optimal circuit parameters  to produce the resource state in the absence of loss are given by the following: writing $z = r \exp{(i\, {\rm arg}(z))}$ we have $\bs{r}=(0.71, 0.67, -0.42)$, ${\rm arg}\,\bs{z} = (-2.07, 0.06, -3.79)$, $\bs{\alpha}=(-0.02, 0.34, 0.02)$, $\bs{\theta}=(-1.57, 0.68, 2.5)$, $\bs{\phi}=(0.53, -4.51, 0.72)$.} 
\label{3mode-loss}
\end{figure*}

In a recent work~\cite{sabapathy2018near}, an optical circuit was presented to generate states of the form
\begin{align}
\label{psia}
    \ket{\psi_a} = \frac{1}{\sqrt{1+5|a|^2/2}} \left[\ket{0}+ia \sqrt{\frac{3}{2}}\ket{1}+ ia \ket{3} \right], 
\end{align}
that used squeezed displaced states, beamsplitters and PNR detectors in two of the three modes. The lossy version of this circuit is presented in Fig.~\ref{3mode-loss}, where various sources of loss at the initial part of the circuit are combined together as a single loss channel $\mathcal{L}(\eta_2)$, and sources of loss at the measurement part of the circuit combined into the loss channel $\mathcal{L}(\eta_1)$. 

In the ideal case, this state can be  produced with perfect fidelity and a probability $> 1\%$. The state is useful since it can be combined with a standard gate teleportation protocol to implement a weak cubic-phase gate $V(\gamma) = \exp [i \gamma \hat{x}^3]$ (see also Table II of Ref. \cite{krishnaON}). It is the lowest-order gate among non-Gaussian quadrature phase gates, but it is still a challenge to implement.
\begin{figure*}
\begin{center}
\scalebox{0.5}{\includegraphics{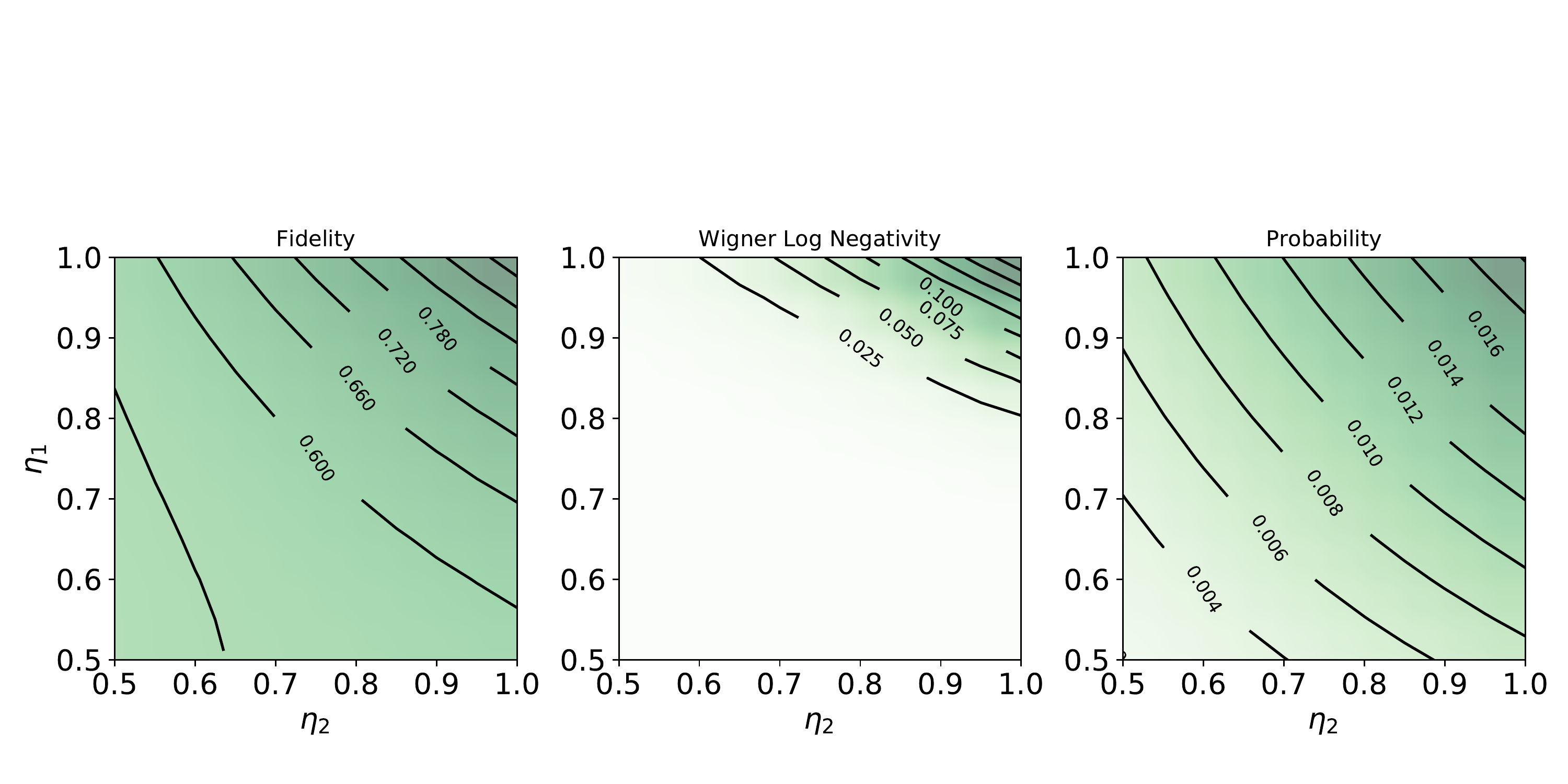}}
\end{center}
\caption{Figures of merit for the weak cubic-phase state preparation scheme as a function of the transmission coefficients in each mode as depicted in Fig.~\ref{3mode-loss}.  The fidelity is with respect to the target state in Eq.~\eqref{psia} with $a=0.53$. Losses must be very low for high fidelity and WLN, although there are regions in the parameter space that correspond to states with non-zero WLN even if its fidelity to the target state is low. The values for WLN and success probability in the absence of loss are given by 0.224 and 0.02 respectively.}
\label{fig:cubic}
\end{figure*}

We now investigate the effect of loss in both modes corresponding to the target state with $a=0.53$ in Eq.~\eqref{psia} when compared with the lossless case. The optimal values of the loss-free circuit parameters used to produce the state can be found in the Supplementary Material of Ref. \cite{sabapathy2018near} and are reproduced for convenience in the caption of Fig.~\ref{3mode-loss}. We plot the fidelity, probability, and WLN in Fig.~\ref{fig:cubic} for $\eta_1,\eta_2 \in [0.5,1]$. We find that if we require the fidelity of the output state with respect to the target state to be larger than $90\%$, then both $\eta_1,\eta_2$ need to be close to unity. On the other hand, we find that genuine non-Gaussianity is still present for states with non-unit transmission coefficients. This observation provides motivation for the need for useful distillation protocols to convert the noisy output states to high quality target states. We find that the state is more fragile with regard to $\eta_1$ when compared to $\eta_2$. 

\section{Conclusion \label{conc}}
Non-Gaussian states are a resource for several useful quantum information processing tasks, and therefore their high-quality preparation  is a crucial goal for experimental quantum optics. One of the major obstacles in this endeavour is the presence of imperfections, whose role must first be understood before their effects can be mitigated or overcome. However, the analysis and simulation of conditional state preparation schemes using practical imperfections in multimode circuits can be computationally challenging.

The simulation algorithm presented in this work is designed to alleviate these hurdles, presenting a faster method to study conditional multimode state preparation schemes in the presence of pure-loss channels. This toolbox may prove useful for both theorists and experimentalists aiming to design new circuits to generate photonic resource states. By applying the simulation algorithm to several illustrative state preparation examples, we confirm that it is indeed challenging to recover large fidelities and Wigner logarithmic negativities in the presence of photon loss. This points to a requirement to explore more sophisticated state preparation schemes that may be able to circumvent these limitations, and we expect our simulation strategy to help in this direction. Looking forward, our simulation method can be paired with optimization tools to find optimal output states given a circuit architecture. Also other experimental imperfections such as dark counts can be easily accounted in our simulation method by replacing loss channels by thermal loss channels before detection.  \\ 

\noindent he code used to generate the results presented here can be found at \cite{code}. \\

\noindent
{\em \bf Acknowledgements}\,: We thank Dylan Mahler, Haoyu Qi, Maria Schuld, Leonardo Banchi, and Daiqin Su for valuable  comments. 

\appendix

\section{Fock matrix elements of Gaussian states}\label{app:fockgaussian}

To obtain the matrix elements of a Gaussian state in the Fock basis we follow the same strategy of Refs.~\cite{kruse2018detailed,hamilton2017gaussian}, i.e., we write the $Q$ function \cite{husimi1940some} of a Gaussian state $\rho$ and the $P$ function of the operator $\ket{\bm{n}} \bra{\bm{m}}$, where $\ket{\bm{n}} \ (\bra{\bm{m}})$ are multimode Fock state kets (bras), and perform a phase space integral as will be demonstrated shortly. 
It generalizes the results from Refs.~\cite{kruse2018detailed,hamilton2017gaussian} as well as those for single-mode systems from Refs.~\cite{marian1993squeezed,marian1993squeezed2}, in that it now allows for $\bm{n} \neq \bm{m}$. It further generalizes the results of Ref. \cite{quesada2018faster} in that one can also calculate matrix elements of Gaussian \emph{mixed} states with finite means.

The $Q$ function of an $\ell$-mode Gaussian state $\rho$ is\,\cite{hamilton2017gaussian}
\begin{align}
	Q_\rho(\vec{\alpha}) = \frac{1}{\sqrt{\det(\pi \bm{\sigma}_Q)}} \exp\left(-\frac{1}{2} (\vec \alpha - \vec \beta)^\dagger \bm{\sigma}_Q^{-1} (\vec \alpha - \vec \beta) \right),
\end{align}
where 
$\bm{\sigma}_Q = \bm{\sigma} +\tfrac{1}{2} \mathbb{I}_{2\ell}$, $\bm{\sigma}$ is the Wigner covariance matrix defined in Eq.  \eqref{eq:WignerCovMat}, $\vec \beta$ is the vector of means defined in Eq.  \eqref{eq:WignerMeans} and we define $\vec \alpha^T = (\alpha_1,\alpha_2,\ldots, \alpha_\ell, \alpha_1^*,\alpha_2^*,\ldots, \alpha_\ell^*)$.

We are interested in the following matrix element 
\begin{subequations}
\label{probn}
\begin{align}
\langle \bm{m}&|\rho|\bm{n} \rangle \nonumber \\
&= \pi^\ell \int d  \vec{\alpha} \ P_{\ket{\bm{n}} \bra{\bm{m}}}(\vec{\alpha}) \  Q_\rho(\vec{\alpha}) \label{probn:a}
\\
&=   \int d  \vec{\alpha} \prod_{s=1}^\ell \frac{\exp\left(|\alpha_s|^2 \right)}{\sqrt{n_s! m_s!}} \left( \partial_{\alpha_s}^{n_s} \partial_{\alpha_s^*}^{m_s} \delta(\alpha_s)  \delta(\alpha_s^*) \right)   \label{probn:b}
\\
& \quad \times \frac{ (-1)^{n_s+m_s} }{\sqrt{\text{det}(\bm{\sigma}_Q)}}\exp\left(-\frac{1}{2} (\vec{\alpha}-\vec{\beta})^\dagger \bm{\sigma}_Q^{-1} (\vec{\alpha}-\vec{\beta}) \right)  \nonumber  \\
&=  \frac{1}{\sqrt{\text{det}(\bm{\sigma}_Q)}} \prod_{s=1}^\ell  \frac{\partial_{\alpha_s}^{n_s} \partial_{\alpha_s^*}^{m_s}}{\sqrt{n_s! m_s!}} \label{probn:c}
\\
& \quad \times \left. \exp\left( \frac{\vec{\alpha}^\dagger \vec{\alpha}}{2} -\frac{(\vec{\alpha}-\vec{\beta})^\dagger \bm{\sigma}_Q^{-1} (\vec{\alpha}-\vec{\beta})}{2}   \right) \right|_{\vec \alpha=0}, \nonumber
\end{align}
\end{subequations}
where the vectors $\bm{m} = (m_1,\ldots,m_\ell)$ and $\bm{n} = (n_1,\ldots,n_\ell)$ contain non-negative integers and we have used the expression for the $P$ function of the operator $\ket{n} \bra{m}$ derived in Appendix \ref{sec:pnm} to write Eq. (\ref{probn:b}). In going from Eq. (\ref{probn:b}) to Eq. (\ref{probn:c}) we used the multidimensional extension of the well-known property of the derivative of a delta function $\int d\alpha d\alpha^* f(\alpha,\alpha^*)\partial_{\alpha}^n \delta(\alpha) \partial_{\alpha^*}^m \delta(\alpha^*) = (-1)^{n+m} \partial_{\alpha}^n \partial_{\alpha^*}^m f(\alpha,\alpha^*)$.

The argument inside the exponential of Eq. (\ref{probn:c}) can be rearranged as follows
\begin{align} 
	&\vec{\alpha}^\dagger \vec{\alpha}- (\vec{\alpha}-\vec{\beta})^\dagger (\bm{\sigma}_Q)^{-1} (\vec{\alpha}-\vec{\beta})   \\
	&= \vec{\alpha}^\dagger (\mathbbm{1}_{2\ell} - \bm{\sigma}_Q^{-1}) \vec{\alpha}  -  \vec{\beta}^\dagger \bm{\sigma}_Q^{-1} \vec{\beta} 
+  \vec{\alpha}^\dagger  \bm{\sigma}_Q^{-1} \vec{\beta}+  \vec{\beta}^\dagger  \bm{\sigma}_Q^{-1} \vec{\alpha}. \nonumber
\end{align}
Note that the (hermitian-)matrix $\bm{\sigma}_Q$ and its inverse have the block structure  $\left[	\begin{smallmatrix}
		\bm{W} & \bm{Y}^* \\
		\bm{Y} & \bm{W}^* \\
	\end{smallmatrix} \right],
$
where  $\bm{W} = \bm{W} ^\dagger  \in \mathbb{C}^{\ell \times \ell}$ is Hermitian  and $\bm{Y} = \bm{Y} ^T \in \mathbb{C}^{\ell \times \ell}$ is symmetric. Motivated by this observation we define the following symmetric matrices
\begin{align}
\bm{A} &= \bm{X}	\left(\mathbbm{1}_{2\ell} - \bm{\sigma}_Q^{-1} \right), \quad \bm{X} =  \left[\begin{smallmatrix}
	0 &  \mathbbm{1}_\ell \\
	\mathbbm{1}_\ell & 0  
\end{smallmatrix} \right],
\end{align}
and further simplify the linear terms in $\vec \alpha$ as
\begin{align}
  \vec{\alpha}^\dagger  \bm{\sigma}_Q^{-1} \vec{\beta} =   \vec{\beta}^\dagger  \bm{\sigma}_Q^{-1} \vec{\alpha} =   \vec \gamma^T  \vec \alpha, \text{ with }  \vec \gamma^T =   \vec{\beta}^\dagger \bm{\sigma}_Q^{-1}.
\end{align}
We obtain these simplification due to the block structure of $\bm{\sigma}_Q^{-1}$ and $\vec \alpha$; for these same reasons the vector $\vec \gamma$ also has a block structure where the second half of the components are the complex conjugate of the first half.

We can now write the density matrix element of interest as
\begin{align}\label{eq:simple}
\langle \bm{m}|\rho|\bm{n} \rangle  &= T  \prod_{s=1}^\ell  \partial_{\alpha_s}^{n_s} \partial_{\alpha_s^*}^{m_s} \left. \exp\Big(\underbrace{\tfrac{1}{2} \vec \alpha^T \bm{A} \vec \alpha +  \vec \gamma^T  \vec \alpha}_{\equiv g(\vec \alpha)} \Big) \right|_{\vec \alpha=0},
\end{align}
where we have collected all the prefactors into
\begin{align}
T =  \frac{\exp\left(-\tfrac{1}{2} \vec \beta^\dagger \bm{\sigma}_Q^{-1} \vec \beta \right)}{ \sqrt{\text{det}(\bm{\sigma}_Q) \prod_{s=1}^\ell n_s! m_s!} },
\end{align}
and used $\vec \alpha^\dagger \bm{X} = \vec \alpha^T $ and $\bm{X}^2 = \mathbbm{1}_{2 \ell}$.

We now need to evaluate the derivatives appearing in Eq.~\eqref{eq:simple}. Let us first consider the case where $0 \leq n_s,m_s \leq 1$; the general case is considered at the end of this section. Using Eq.~\eqref{eq:simple} and Fa\`a di Bruno's generalization of the chain rule for partial derivatives (cf. Eq.~(4) of~\cite{hardy06}) for an exponential function $\exp\left(g(\vec{\alpha}) \right)$ we find
\begin{align}
\label{hardy}
&\partial_{\vec \alpha_{i_1} } \cdots \partial_{\vec \alpha_{i_K}}  \exp\left(g(\vec{\alpha})\right) \\
& = \exp\left( g(\vec{\alpha}) \right) \sum_{\pi \in \mathcal{P}[\{ {i_1},\ldots, {i_K}\}]} \prod_{B\in \pi} \left( \prod_{j \in B} \partial_{\vec \alpha_{j}}   g(\vec{\alpha}) \right), \nonumber
\end{align}
where $ \{i_1,...,i_K\} =I$ is an index set, the sum over $\pi$ runs over all partitions $\mathcal{P}[\{ {i_1},\ldots, {i_K}\}] $ of this set of indices, and $B$ runs over the ``blocks'' from that partition. For example, if the set of indices is $\{1,2,3,4\}$ then the partitions are
\begin{align}
\pi &= \{ \{1,2,3,4\}, \{1\} \{2,3,4\}, \{2\} \{1,3,4\}\\
&,\ldots , \{1,2\}\{3,4\},\ldots,  \{1\} \{2\}\{3\}\{4\}  \} \nonumber.
\end{align}
The blocks of the partition $\{1\} \{2,3,4\}$ are $\{1\}$ and $\{2,3,4\}$.\\

For our problem, the vectors $\bm{n}$ and $\bm{m}$ are what determines the index $I$. In particular recall that 
\begin{align}
	\alpha_s  = \vec \alpha_s \text{ but, } \alpha_s^* = \vec \alpha_{s+\ell},	
\end{align}
thus if $n_s=1 \ (n_s=0)$ then $s \in I \ (s \not\in I)$, similarly if $m_s=1 \ (m_s = 0)$ then $s+\ell \in I \ (s+\ell \not \in I)$. Using the fact that $g(\vec \alpha)$ is quadratic in $\vec \alpha$ we find
\begin{align}
 \partial_{\vec \alpha_k}  g(\vec{\alpha}) &=  \vec{\gamma}_k, \label{zero_disp}\\
\partial_{\vec \alpha_s} \partial_{\vec \alpha_k}  g(\vec{\alpha}) &= A_{s,k},
\end{align}
and that all higher order derivatives are zero. This means that only first and second order derivatives survive and the sum over partitions of the index set $\pi$ collapses to a sum where only partitions with subsets/blocks of at most size $2$ survive. If we label the vertices of a graph with the index set $I$ then the sum can also be understood as going over the single-pair matchings (SPMs) of the graph~\cite{quesada2018faster,bjorklund2018faster}. This is the set that enumerates the perfect matchings of a graph \emph{with loops}. For the index set $I$ we write it as SPM$(I)$. 

Whenever $n_s=0 \ (m_s=0)$ we are instructed to not include the integer $s \ (s+\ell)$ in the index set $I$. Thus we can construct the matrix $\bar{\bm{A}}$ obtained from $\bm{A}$ by keeping only the rows and columns labelled by the elements of $I$. Similarly we can construct $\bar{\gamma}$ from $\vec \gamma$ by keeping only the components indexed by the elements of $I$.
With this notation we can finally write 
\begin{align}
\bra{\bm{m}}\rho \ket{\bm{n}} &=  T  \prod_{s=1}^\ell  \partial_{\alpha_s}^{n_s} \partial_{\alpha_s^*}^{m_s} \left. \exp\left(\tfrac{1}{2} \vec \alpha^T \bm{A} \vec \alpha +  \vec \gamma^T  \vec \alpha \right) \right|_{\vec \alpha=0} \nonumber \\
 &=   T \sum_{\pi \in \text{SPM}(I) }	\prod_{(i,j) \in \pi } \tilde A_{i,j} \nonumber\\
 &= T \times  \text{lhaf}(\tilde{\bm{A}}), 
\end{align}
where we have defined
\begin{align}
	\tilde A_{i,j} = \begin{cases}
		\bar{A}_{i,j} &\text{ if } i\neq j,\\
		\bar{\gamma}_{i} &\text{ if } i=j,
	\end{cases}
\end{align}
and lhaf is the loop hafnian \cite{bjorklund2018faster}. Note that whenever $\bm{n} = \bm{m}$ and $\vec{\beta} = \vec \gamma = 0$ the above result reduces to the hafnian \emph{probabilities} calculated by Hamilton \emph{et al.} \cite{hamilton2017gaussian}.

Let us now consider the multiphoton case where $n_s, m_s > 1$. As was done earlier in Refs.  \cite{hamilton2017gaussian,kruse2018detailed,bradler2018graph,quesada2018faster}, we state that this case can be dealt with by introducing auxiliary modes in which $n_s$ photons in a single mode are mapped to $n_s$ single photons in $n_s$ modes. This implies that whenever $n_s$ (or $m_s$) is greater than one we simply need to repeat the value $s$ (or $s+\ell$) in the index $I$ a total of $n_s$ (or $m_s$) times.

\section{The Glauber-Sudarshan function of $\ket{n}\bra{m}$}\label{sec:pnm}
Before starting with the derivation we state two useful identities. The first one was derived by Mehta \cite{mehta67diagonal} and gives the $P$ function of any operator $\rho$ as follows
\begin{align}\label{mehta}
	P_\rho(\alpha) =  e^{|\alpha|^2} \int \frac{d^2 \beta}{\pi^2} \ \langle -\beta |\rho|\beta \rangle  \	e^{|\beta|^2} \ e^{\beta^* \alpha - \beta \alpha^*}.
\end{align}
The second is simply a resolution of the identity 
\begin{align}\label{identity}
	\int \frac{d^2 \beta }{\pi^2}	e^{\beta^* \alpha - \beta \alpha^*} = \delta(\alpha) \delta(\alpha^*).
\end{align}
Now we set $\rho = \ket{n}\bra{m}$ in Eq. \eqref{mehta} and, using
\begin{align}
	\langle -\beta|n \rangle = e^{-|\beta|^2/2} \frac{(-\beta^*)^n}{\sqrt{n!}}, \quad
	\langle m|\beta \rangle = e^{-|\beta|^2/2} \frac{(\beta)^m}{\sqrt{m!}},
\end{align}
find
\begin{align}
	P_{\ket{n}\bra{m}}(\alpha) 	 = e^{|\alpha|^2} \int \frac{d^2 \beta}{\pi^2} \ \frac{(-\beta^*)^n (\beta)^m }{\sqrt{m! n!}} \ e^{\beta^* \alpha - \beta \alpha^*}.
\end{align}
To make progress first consider the case $n=m=0$ for which we find
\begin{align}
	P_{\ket{0}\bra{0}}(\alpha) 	 = e^{|\alpha|^2} \int \frac{d^2 \beta}{\pi^2} \  e^{\beta^* \alpha - \beta \alpha^*} = \delta(\alpha) \delta(\alpha^*),
\end{align}
where we used the resolution of the identity of Eq. (\ref{identity}). Now note the following
\begin{align}
	(-\beta^*)^n (\beta)^m   e^{\beta^* \alpha - \beta \alpha^*}	 = (-1)^{n+m}  \partial^n_{\alpha} \partial^m_{\alpha^{*}} e^{\beta^* \alpha - \beta \alpha^*}.
\end{align}
Using this identity we find
\begin{align}
	P_{\ket{n}\bra{m}}(\alpha) 	 =  \frac{e^{|\alpha|^2}}{\sqrt{m! n!}}	(-1)^{n+m}  \partial^n_{\alpha} \partial^m_{\alpha^{*}} \delta(\alpha) \delta(\alpha^*),
\end{align}
which reduces to the well known result in Eq. 4.4.51 of Gardiner and Zoller \cite{gardiner2004quantum} when $n=m$. Also note that the $P$ function of $\ket{n}\bra{m}$ was derived in polar coordinates by Sudarshan \cite{sudarshan1963equivalence}; we just provide the derivation  in the $\alpha,\alpha^*$ basis for completeness.

\bibliographystyle{apsrev}
\bibliography{biblio}

\end{document}